\documentclass[]{raa}            
\usepackage{graphicx,times}
\usepackage{natbib}

\providecommand{\bm}[1]{\mbox{\boldmath$#1$\unboldmath}}

\begin{document}

   \title{On Pulsar-Driven Mass Ejection in Low-Mass X-ray Binaries}

 \volnopage{ {\bf 0000} Vol.\ {\bf 0} No. {\bf 00}, 000--000}
   \setcounter{page}{1}

\author{Lei Fu
\and Xiang-Dong Li
   }
   \institute{Department of Astronomy, Nanjing University, Nanjing 210093, China;
   {\it fl19821110@yahoo.com.cn; lixd@nju.edu.cn}\\
   Key Laboratory of Modern Astronomy and Astrophysics,
   Ministry of Education, Nanjing 210093, China
\\
\vs \no
   {\small Received [year] [month] [day]; accepted [year] [month] [day] }
} \abstract{There is accumulating evidence for mass ejection in
low-mass X-ray binaries (LMXBs) driven by radio pulsar
activity during X-ray quiescence. In this paper we consider the
condition for mass ejection by comparing the radiation
pressure from a millisecond pulsar, and the gas pressure at the
inner Lagrange point or at the surrounding accretion disk. We
calculate the critical spin period of the pulsar below which mass
ejection is allowed. Combining with the evolution of the mass transfer
rate, we present constraints on the orbital periods of the systems. We
show that mass ejection could happen in both wide and compact LMXBs.
It may be caused by transient accretion due to thermal instability in the
accretion disks in the former, and irradiation-driven mass-transfer cycles
in the latter.
 }

   \authorrunning{Fu \& Li}
   \titlerunning{Pulsar-driven mass ejection}
   \maketitle

\section{Introduction}
\label{sect:intro}

Currently there are 144  millisecond pulsars (MSPs) (pulse periods
$p<10$ ms) in the ATNF pulsar database, 86 of which are located in
binary systems. MSPs  are traditionally considered as the descendants of old
neutron stars which were spun up by accretion from their companions in
low-mass X-ray binaries (LMXBs)  (Bhattacharya
\& van den Heuvel 1991 for a review). Typically, accretion of $\Delta
M \sim 0.1 M_\odot$ mass is sufficient to accelerate a slowly
rotating neutron star to milliseconds (Burderi et al. 1999).

 More recent investigations demonstrate that perhaps most of the current LMXBs have
evolved from systems with intermediate-mass ($\ga 1.5\,M_{\odot}$)
donor stars, i.e., intermediate-mass X-ray binaries (IMXBs) (Davies
\& Hansen 1998; King \& Ritter 1999; Podsiadlowski et al. 2002;
Pfahl et al. 2003). Because the donor star is initially more massive
than the neutron star, mass transfer proceeds rapidly on a timescale
$\la 10^6$ years. Little mass is accreted during this
phase since the mass transfer rate is usually much higher than the
Eddington limit for the neutron star, until the donor mass becomes
comparable with the neutron star mass and the binary evolves to be
an LMXB (Tauris \& Savonije 1999; Podsiadlowski et al. 2002). It is
believed that in this later phase the neutron star accretes
sufficient mass and experiences the recycling process.

There is evidence that mass transfer in the LMXB phase is also
highly nonconservative. The companion star of MSPs is usually a
white dwarf of mass $\sim 0.1-0.4 M_\odot$, the progenitor of which
is expected to be a star with initial mass $\ga 1 M_\odot$
(Tauris \& Savonije 1999; Pfahl et al. 2003). So the lost mass from
the companion is about $0.6-0.9M_\odot$. If the mass transfer is
conservative, the expected minimum mass of the MSPs  would exceed
$2M_\odot$. However, measurements of the pulsar masses suggest that
only part of the transferred mass was accreted by the neutron stars
(Thorsett \& Chakrabarty 1999; Zhang et al. 2011: Kizilatan, Kottas,
\& Thorsett 2010. See also Table 1 for a list of binary MSPs with
measured masses), except that  in a few case the masses of MSPs are
as high as $\sim 2M_\odot$ (e.g., PSRs B1957+20, van Kerkwijk \&
Breton 2011; J0751+1807, Nice et al. 2005; J1614-2230, Demorest et
al. 2010\footnote{In a recent paper, Tauris et al. (2011) suggested
that PSR J1614$-$2230 must have been born with a mass significantly
exceeds $1.4\,M_{\odot}$.}). So in most of the systems, the mass
transfer must have been nonconservative.

The mechanism of mass ejection in X-ray binaries was first
introduced by Illarionov \& Sunyaev(1975) as the ``propeller"
effect, i.e., the accretion flow is centrifugally prohibited at the
magnetosphere of the neutron star, at the cost of the slow down of
star's spin. As pointed out by Burderi et al. (2001), the efficiency
of the propeller effect is at most $\sim 50\%$, and the pulsars
should still be more massive than observed. It was later suggested
that, if the mass transfer rate varies by large amplitude, the neutron star may
become an MSP when the accretion rate is very low, and
the energetic pulsar wind may disrupt the accretion
disk around the neutron star, so that the transferred mass will
escape from the binary across
 the inner Lagrangian point $L_1$, which we call pulsar-driven mass ejection
(Burderi et al. 2001; Burderi, D'Antona \& Burgay 2002). Since this
process occurs at $L_1$, at which the binding energe is very small,
the efficiency of ejeciton may reach unity. An alternative mechanism
is that the energetic wind or hard X-ray radiation from the MSP may
evaporate it's low-mass companion (van den Heuvel \& van Paradijs
1988; Ruderman, Shaham \& Tavani 1989;
Shaham \& Tavani 1990;  Shaham \& Tavani 1991; Podsiadlowski 1991).

Direct observational evidence for mass ejection induced by pulsar
activity may come from the ``black widow pulsars". They are MSPs in
binary systems, undergoing very wide eclipse, implying obscuration
by intense wind from the secondary (Fruchter, Stinebring, \& Taylor
1988). Another interesting example is PSR J1740$-$5340, an eclipsing MSP
with a spin period of 3.65 ms and orbital period of 32.5 hr, located
in the globular cluster NGC 6397 (D'Amico et al. 2001). Long lasting
and sometimes irregular radio eclipses, and the shape of the optical
light curve demonstrated the presence of matter around the system
(Ferraro et al. 2001),  suggesting that PSR J1740$-$5340 is an
example of a system in the pulsar-induced ejection phase.

In this paper we investigated the conditions for mass ejection in
LMXBs caused by the turn-on of pulsar activity. The basic idea is
similar to that in Burderi et al. (2001). However, we consider
direct mass ejection at both $L_1$ and $L_2$, which requires that
the pulsar's radiation pressure is strong enough to drive material
out of these points, and was ignored in Burderi et al. (2001). For
mass ejection due to disruption of the accretion disks by the
pulsar's pressure, we adopt a more realistic accretion disk model to
evaluate the disk pressure. In section 2 we calculated the critical
periods $p_{\rm cr}$ at which the pulsar wind pressure $P_{\rm psr}$
equals the donor's gas pressure $P_{\rm L1}$  at the $L_1$ point,
and $P_{\rm disk}$ in the accretion disk respectively. The condition
for outflow from $L_2$ is also considered. In section 3 we put
possible constraints on the orbital periods by combining the
evolution of LMXBs. We briefly discuss the implications of our
results in section 4.

\section{The Critical Periods}
\label{sect:CPs}

\subsection{Mass Flow Through the Inner Lagrange Point $\bm L_1$}

We consider a semi-detached binary containing a neutron star and a
donor star that fills it's Roche lobe. The atmospheric material is
overflowing through $L_1$  towards the neutron star. At $L_1$  the
gas flow is confined in a ``nozzle" with radius $H$ and moves with a
speed close to the sound velocity (Frank, King, \& Raine 2002). The
mass transfer rate is expressed as
\begin{equation}
\dot{M} \simeq \pi H^2 {c_s}(L_1) \rho (L_1),
\end{equation}
where ${c_s}(L_1)$ and $\rho (L_1)$ are the local sound velocity and
the gas density at the $L_1$ point. To estimate the magnitude of
$H$, we consider the balance between the Roche potential and the
mean kinetic energy of the material, and find that matter escapes in
a patch of radius (see also Frank, King, \& Raine 2002; Li et al.
2010)
\begin{equation}
H \simeq {{c_s}(L_1)\over{\sqrt{f_1} \omega}},
\end{equation}
where
\begin{equation}
{f_1}(q)={1 \over 2} \left[ {{1} \over
{(1+q)(0.5-0.227\log_{10}q)^3}} + {{q} \over
{(1+q)(0.5+0.227\log_{10}q)^3}} -1 \right],
\end{equation}
where $q$ is the mass ratio of the donor and the neutron star,
ranging from $\sim 0.1$ to $\sim 3$ for I/LMXBs. The mass transfer
rate can then be written as
\begin{equation}
\dot{M} = 1.03 \times 10^6 P_{\rm orb,hr}^{2} c_{s}^{3}(L_1) \rho(L_1) f_{1}^{-1}(q) {\rm g\,s}^{-1},
\end{equation}
where $P_{\rm orb,hr}$ is the orbit period of the system in units of
hour. Consequently, the gas pressure at $L_1$ is
\begin{equation}
P_{L1} = \rho(L_1)c_{s}^2(L_1)=6.15 \times 10^3\dot{M}_{-10} P_{\rm
orb,hr}^{-2} c_{s6}^{-1}f_{1}\, {\rm dyn\,cm}^{-2},
\end{equation}
where $\dot{M}_{-10} =\dot{M}/10^{-10}\,M_{\odot}$yr$^{-1}$, and
$c_{s6}=c_s/10^6$\,cms$^{-1}$.

\subsection{Equilibrium at $\bm L_1$ and $\bm L_2$}

Once the material from the donor star is captured by the neutron
star at the circularization radius, an accretion disk will form due
to dissipative processes and internal torque in it. Sufficiently
long time accretion on to the neutron star can spin it up to
millisecond periods. If the accretion rate drops due to some
reasons, the inner radius of the accretion disk may move outside of
the light cylinder (with radius $R_{LC}$) of the neutron star, and
the neutron star changes from an accreting X-ray source to be a
rotation-powered pulsar, emitting out radio, optical, X-ray,
$\gamma$-ray photons and high energy particles. If this process is
dominated by the magnetic dipole radiation, the pulsar wind pressure
at $L_1$ can be expressed as
\begin{equation}
P_{\rm psr}(L_1)=8.23 \times 10^2 B_{s8}^2 P_{\rm
orb,hr}^{-4/3}p_{\rm ms}^{-4}R_{n6}^{6}m_1^{-2/3}f_2^{-2}(q) {\rm
\,dyn\,cm}^{-2},
\end{equation}
where $m_1$ and $R_{n6}$ are the mass and radius of the pulsar in
units of solar mass and $10^6$ cm respectively, $B_{s8}$ the surface
magnetic field in units of $10^8$ G, $p_{\rm ms}$ the pulsar period
in units of millisecond, and
\begin{equation}
f_2(q)=(1+q)^{1/3}(0.5-0.227\log_{10}q).
\end{equation}
When the pulsar wind pressure is larger than the gas pressure
($P_{L1}$) at $L_1$, the accretion process is prohibited, and mass
ejection will happen. Combining Eqs.~(5) and (6) we can get a
critical period at which material flow through $L_1$ will be
ejected,
\begin{equation}
p_{\rm cr}\simeq (0.61\,{\rm ms})~
f_1^{-1/4}f_2^{-1/2}\dot{M}_{-10}^{-1/4}
c_{s6}^{1/4}B_{s8}^{1/2}P_{\rm orb,hr}^{1/6}R_{n6}^{3/2}m^{-1/6}.
\end{equation}

When the secondary star evolves to be of very low mass, it may
be the outer Lagrangian point $L_2$ rather $L_1$ with the lowest
potential, due to the pulsar's radiation pressure force on the Roche
potential of the secondary. In this case, Roche-lobe overflow will
not occur towards the pulsar, but into a circumbinary disk/outflow
(as is indeed observed for PSR 1957$+$20). According to Phillips \&
Podsiadlowski (2002), the condition for outflow from $L_2$ can be
defined as $\delta_{max}\equiv qL/L_{Edd}$ becomes larger than a
critical value $\delta_{crit}$ (where $L$ and $L_{Edd}$ are the
irradiating pulsar's luminosity and the Eddington luminosity of the
secondary, respectively), or equivalently
\begin{equation}
{{L \kappa} \over {4\pi c GM}}=\delta_{crit},
\end{equation}
where
\begin{equation}
L=\left({2R_n^6} \over {3c^3} \right)B_s^2 \left({2\pi}\over
p\right)^4,
\end{equation}
and $\kappa$ is the mean photospheric opacity. The resulting
critical spin period is
\begin{equation}
p_{\rm cr, L2}=2\pi \left({{\kappa B^2} \over {4\pi cGM
\delta_{crit}}}{{2R_n^6} \over {3c^3}}\right)^{1/4}
\end{equation}
or
\begin{equation}
p_{\rm cr, L2}\simeq (0.48\,{\rm ms})~ \kappa_{0.4}^{1/4} m^{-1/4}
B_{s8}^{1/2} R_{n6}^{3/2}
\end{equation}
with $\delta_{crit}=0.0564$ when $q=0.1$ (Phillips \& Podsiadlowski
2002). Here $\kappa_{0.4}=\kappa/0.4$ cm$^2$g$^{-1}$ may be much
larger than unity for the photosphere of very low-mass stars, so
Eq.~(12) presents a lower limit of the critical period. Obviously,
depending on the mass ratio, this criterion may be more relevant for
very low-mass binaries.

\subsection{Equilibrium with the disc}

Now we consider possible disruption of the accretion disk by the
pulsar's pressure. The standard geometrically thin $\alpha$-disk
model was established by Shakura \& Sunyaev (1973). However, there
is observational evidence showing deviation from the theoretical
predictions -- in cataclysmic variables (CVs) and LMXBs the outer
disk edge seems to be thick and structured (Shafter \& Misselt 2006;
Hakala et al. 1999). Begelman \& Pringle (2007) argued that, the
magnetorotational instability (MRI), which successfully accounts for
angular momentum transport in accretion disks (Balbus \& Hawley
1998), can amplify the toroidal magnetic field to a point at which
magnetic pressure far exceeds the combined gas and radiation
pressure in the disk. This additional pressure support makes the
disk thicker than in the $\alpha$-disk model, in accordance with
observations. In this magnetic-dominated disk model, the Alfv\'en
speed associated with the toroidal field is roughly the geometric
mean between the Keplerian speed and the sound speed. Thus the
magnetic pressure is $P_B \simeq \rho c_{g} v_k$. The total disc
pressure can be derived to be (see appendix)
\begin{equation}
P_{\rm disk} \simeq P_B=1.02 \times 10^{13} \alpha^{-17/18}m^{61/36}
\dot{M}_{-10}^{8/9}R_6^{-91/36} R_{n6}^{-1} \left[ 1-{\left ({R_{n6}
 \over R_6} \right)}^{1/2} \right]^{-1/9}\,{\rm dyn\,cm}^{-2},
\end{equation}
where $R_6$ is the disk radius in unit of $10^6$ cm. As pointed out
by Burderi et al. (2001), if the spun-up pulsar is switched on, the
accretion disk may be truncated at a radius $R_{\rm stop}$ due to
the energetic pulsar wind. Once $R_{\rm stop}$ is smaller than the
outer radius $R_{\rm out}$ of the accretion disk, the overflowed
material through $L_1$ will be ejected, i.e., the system will enter
the mass ejection phase.

Traditionally the outer radius of the disk is approximated to be
$\sim 70\%-90\%$ of the Roche lobe radius $R_{L1}$ of the primary
(Frank, King, \& Raine 2002). Here we take $R_{\rm out} \simeq
0.8R_{L1}$. Combining Eq.~(6) and (13) , the critical pulsar spin
period for mass ejection is
 \begin{equation}
 p'_{\rm cr}=(1.94\,{\rm ms})\,\alpha^{17/72} m^{1/24} R_{n6}^{7/4} B_{s8}^{1/2} \dot{M}_{-10}^{-2/9} P_{\rm orb,hr}^{19/216} f_3
 \end{equation}
where
\begin{displaymath}
f_3=(1+q)^{19/432}\left[1-0.462 \left( {q} \over {1+q}\right)^{1/3}\right]^{91/144}(0.5-0.227\log_{10}q)^{-1/2}.
\end{displaymath}

\subsection{Accretion Equilibrium Period}
In the above subsections we show that mass ejection can take place
only when the spin period of the neutron star is shorter than either
$p_{\rm cr}$ or $p'_{\rm cr}$. The neutron star's spin is
accelerated by both mass accretion and magnetic field-disk
interaction during the mass transfer process, until an equilibrium
period is reached, at which there is no net torque exerted on the
neutron star. According to Ghosh \& Lamb (1979), the spin-up/down
rate depends on the fastness parameter $\omega_{\rm
s}\equiv\Omega_{\rm s}/\Omega_{\rm K}(r_0)$, i.e., the ratio of the
spin angular velocity of the neutron star and the Keplerian velocity
of plasma at the inner disk radius $r_0$. When  $\omega_{\rm s}$
approaches the critical fastness parameter $\omega_{\rm c}$, the
torque is zero and the spin-up process ceases. Assume $r_0=0.5R_{\rm
A}$, where $R_{\rm A}$ is the traditional Alfv\'en radius for
spherical accretion
\begin{displaymath}
R_{\rm A}=\left({B_s^2R_n^6} \over {\dot{M}\sqrt{2GM}}\right)^{2/7},
\end{displaymath}
the equilibrium spin period is
\begin{equation}
p_{\rm eq}\simeq (1.95\,{\rm ms})~\omega_{0.7}^{-1}R_{n6}^{18/7}B_{s8}^{6/7}m^{-5/7}\dot{M}_{-10}^{-3/7},
\end{equation}
where $\omega_{0.7}=\omega_{\rm c}/0.7$. The condition $p_{\rm
eq}\le p_{\rm cr}$ or $p'_{\rm cr}$ will constrain the systemic
parameters that allow mass ejection to happen in LMXBs.

The neutron star will reach its equilibrium only when the mass
transfer rate keeps constant for sufficient long time. When the mass
transfer rate changes, the spin period will evolve to new
equilibrium on a time scale $t_{\rm spin}\simeq 2\pi I
\nu_s/\dot{M}(GMR)^{1/2}\sim 2.3\times 10^{9}
p_{ms}^{-1}\dot{M}_{-10}^{-1}$, where $I$ is the momentum of inertia
of the pulsar. Roughly speaking, if the timescale of the change of
the mass transfer rate $t_{\dot{M}}
> t_{\rm spin}$, the spin period of the neutron star is always close to the
equilibrium period with current $\dot{M}$. If $\dot{M}$ varies with
$t_{\dot{M}} \ll t_{\rm spin}$, the spin period remains to be close
to the previous equilibrium value.

\section{Binary MSP Systems that favor material ejection}

\subsection{Mass Transfer Mechanisms}

In this section we investigate the possible influence of the
mass transfer rate. As mentioned above, recent studies showed that
most LMXBs are likely to originate from IMXBs. During the initial
`IMXB to LMXB' phase, the mass transfer process proceeds on a
(sub)thermal timescale, the average mass transfer rate is very high
(often super-Eddington for relatively massive companions), so the
mass transfer is highly unconservative and the accreted mass by the
neutron star is very small. Thus the recycling process should mainly
occur in the later LMXB phase. From Eqs.~(8), (12) and (14) one can
see that mass ejection due to pulsar activity can take place only
when the neutron star's spin period is less than a few milliseconds.
This is difficult to achieve except for LMXBs and close IMXBs (with
case A mass transfer); the outcome of wide IMXB evolution is mildly
recycled MSPs with CO WD companions (Tauris 2011). In the former
case the LMXB evolution from IMXBs is similar to that of original
LMXBs (Podsiadlowski et al. 2002). For these reasons we only
consider the evolution of mass transfer rate in LMXBs.

It's well known that mass transfer in LMXBs is driven by nuclear
expansion of the donor and orbital angular momentum loss due to
magnetic braking and gravitational radiation. By comparing the
nuclear evolution time of the donor $t_{\rm nu}$ and the angular
momentum loss timescale $t_{\rm aml}$, we can divide the
evolutionary tracks of LMXBs into the following cases (King, Kolb,
\& Burderi 1996).

Case 1 $t_{\rm nu}\ll t_{\rm aml}$: In this case the donor has
evolved off the main sequence before angular momentum loss shrinks
the orbit sufficiently to cause mass transfer. So the donor must be
a giant or subgiant, with mass of $\ge 0.8 M_{\odot}$ in order to
evolve off main sequence within Hubble time.  It's known that  the
structure of a low-mass giant is determined by the mass of its
helium core (Webbink et al. 1983). The mass transfer rate can be
expressed by the following relation (King 1988),
\begin{equation}
\dot{M}_{-10}\simeq 0.21 P_{\rm orb,hr}^{0.93}m^{1.47}q^{1.47}.
\end{equation}
Substitute Eq.~(16) into Eqs.~(8), (11), and (15), when $p_{\rm
cr}=p_{\rm eq}$ and $p'_{\rm cr}=p_{\rm eq}$ we obtain the minimum
of the orbital period for mass ejection
\begin{equation}
P_{\rm orb,hr}=75.96~\omega_{0.7}^{-2.94}m^{-2.40}q^{-0.78}B_{s8}^{1.08}
f_1^{0.76}f_2^{1.50}c_{s6}^{-0.76}R_{n6}^{3.21},
\end{equation}
and
\begin{equation}
P'_{\rm orb,hr}=3.22~\omega_{0.7}^{-3.45}\alpha^{-0.83}q^{-1.07}
R_{n6}^{2.83}B_{s8}^{1.24}m^{-3.78}f_3^{-3.45}.
\end{equation}
Figure 1 shows the minimum $P_{\rm orb,hr}$ and $P'_{\rm orb,hr}$
against the mass ratio $q$ in the solid and dashed lines,
respectively.  Note that the minimum orbit period that allows direct
mass ejection at $L_1$ is $\sim 5$ days, while mass ejection with
disk disruption can occur in relatively compact systems. The final
product of this kind of evolution is  a wide binary  with an MSP and
a low-mass helium white dwarf.

We also plot $p_{\rm cr}$, $p'_{\rm cr}$ and $p_{\rm eq}$ for
constant orbital period with black, blue and red lines in Fig.~2 as
a function of mass ratio $q$. Note that in the cases of $P_{\rm
orb,hr}=100$ (dashed lines) and 1000 (solid lines), $p'_{\rm cr}$ is
always larger than $p_{\rm cr}$.

 \begin{figure}
 \centering
   \includegraphics[width=7.0cm]{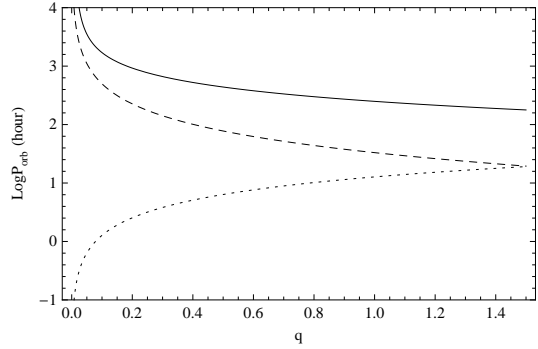}

   \begin{minipage}[]{148mm}

   \caption{ The relation between $P_{\rm orb,hr}$
   and the mass ratio $q$ for LMXBs in case 1 evolution when mass ejection occurs.
   The solid and dashed lines are for $p_{\rm cr}=p_{\rm eq}$,
   and $p'_{\rm cr}=p_{\rm eq}$.
   Above the curves the mass ejection may happen. The dotted line represents the condition
   when the donor fills its Roche lobe on the main sequence. Here and in the following
   we adopt $m=1.4$, $R_{n6}=1$, $B_{s8}=5$ and $\alpha=0.1$. }
   \end{minipage}
   \label{Fig1}
   \end{figure}

\begin{figure}
   \centering
   \includegraphics[width=7.0cm]{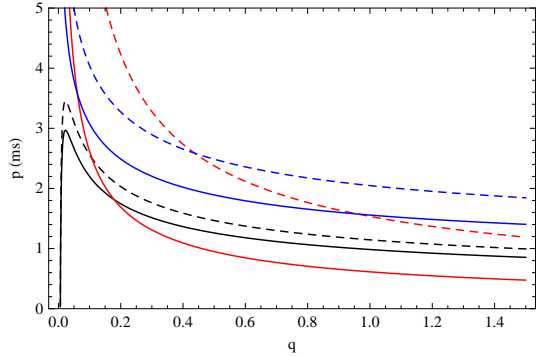}

\begin{minipage}[]{148mm}

   \caption{ The relation between
   critical spin period and the mass ratio $q$ for LMXBs in case 1 evolution.
   The black, blue and red lines describe $p_{\rm cr}$, $p'_{\rm cr}$, and
   $p_{\rm eq}$, respectively. The dashed and solid lines are for
     orbital period $P_{\rm orb,hr} =100, 1000$, respectively. }
   \end{minipage}
   \label{Fig2}
   \end{figure}

Case 2 $t_{\rm nu} \sim t_{\rm aml}$ and $t_{\rm nu}\gg t_{\rm
aml}$: In these situations the mass tranfer is driven by angular
momentum loss of the system, and
\begin{displaymath}
{\dot{M}_{2}\over {M_2}} \sim {\dot{J} \over J},
\end{displaymath}
where $J$ and $\dot{J}$ are the orbital angular momentum and its
derivative, respectively. In the case of $t_{\rm nu}\sim t_{\rm
aml}$ the donor is mildly evolved when it fills it's Roche lobe, and
the radius is slightly bigger than during the main sequence. The
orbit period of the system will not change much during the whole
evolution. When $t_{\rm nu}\gg t_{\rm aml}$, the angular momentum
loss of the system is so rapid that mass transfer occurs when the
secondary is still on the main sequence, and further nuclear
evolution of the secondary is frozen. In this condition the binary
orbit will shrink due to the angular momentum loss by magnetic
breaking and gravitational radiation. Hence the orbital periods are
generally smaller than in case 1 where the secondary is a
(sub)giant.

Use the standard forms of gravitational radiation losses  (Landau \&
Lifschitz 1958) and of magnetic braking (Verbunt \& Zwaan 1981), the
mass transfer rate can be written as (King et al. 1996),
\begin{equation}
\dot{M}_{-10}=552.81P_{\rm orb,hr}^{-2/3}q^{7/3}m^{5/3}+130.65P_{\rm orb,hr}^{-8/3}q^2m^{8/3},
\end{equation}
where the first and second terms on the right hand side are  for
magnetic braking and  gravitational radiation, respectively.
Figure 3 shows  $p_{\rm cr}$, $p'_{\rm cr}$, and $p_{\rm eq}$ as a
function of $P_{\rm orb}$ in black, blue, and red lines, with the
mass transfer rate given by Eq.~(19). The dashed and dotted lines
are for $q= 0.5$ and $0.1$ respectively. In the case of $q=0.1$ we
use Eq.~(19) with donor mass $=0.3\,M_{\odot}$ to calculate $p_{\rm
eq}$, but remove the first term (i.e., corresponding to magnetic
braking) in calculating $p_{\rm cr}$ and $p'_{\rm cr}$. The reason
is that magnetic braking is assumed to vanish abruptly when the
secondary mass decreases to $\sim 0.3\,M_{\odot}$ (i.e., becomes
fully convective), and the subsequent evolution is driven by
gravitational radiation only. However, it will take a long time ($\ga 10^9$ yr) for
the pulsar to reach the new equilibrium period because of the
reduced mass transfer rate. So the spin period remains close to the
previous one. We have set the limitation on the orbital period $
0.11P_{\rm orb,hr}\ge m_2$ to guarantee that the secondary fills its
Roche lobe before it evolves to be a subgiant. Figure 3 shows that
$p_{\rm cr}\la p_{\rm eq}<p'_{\rm cr}$, which means that mass
ejection via disk disruption might occur. But a firm conclusion is difficult
to reach, because of the uncertainties in estimating the critical periods.
By the way, outflow from
$L_2$ seems to require spin periods considerably lower than $p_{\rm eq}$ for
$q=0.1$, disfavoring this possibility for LMXBs. However, this
scenario might work in binaries with extreme mass ratio ($q\ll 1$, like PSR
B1957$+$20), in which the value of $\delta_{crit}$ is much smaller
(Phillips \& Podsiadlowski 2002).

 \begin{figure}
 \centering
  \includegraphics[width=7.0cm]{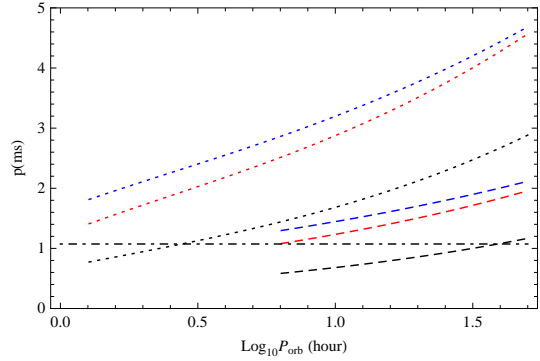}

  \begin{minipage}[]{148mm}

  \caption{This figure compares $p_{\rm cr}$,  $p'_{\rm cr}$, and
   $p_{\rm eq}$ as a function of orbital period in case 2 mass transfer.
   Here $p_{\rm cr}$, $p'_{\rm cr}$, and $p_{\rm eq}$ are plotted in black,
   blue, and red lines, respectively. The dashed and dotted lines are for $q=0.5$
   and 0.1, respectively. The dot-dashed line represents $p_{\rm cr, L2}$ for $q=0.1$.}
   \end{minipage}
   \label{Fig4}
   \end{figure}

Note that $p_{\rm eq}\le p_{\rm cr}$ or $p'_{\rm cr}$ is the
necessary condition for mass ejection, while switch-on of pulsar
emission is critical in expelling the matter overflowing the
Roche lobe. This occurs with a temporary reduction of the mass
accretion rate, so that the accreting plasma moves out beyond the
light cylinder of the neutron star, and the neutron star becomes a
generator of magneto-dipole radiation and relativistic particles.
The thermal disk instability model (Lasota 2001) is usually invoked
to trigger the X-ray outbursts in LMXBs, which may cause the
switch-on of the radio pulsar during X-ray quiescence. According to King et al. (1996),
long-period LMXBs are likely to be transient; for LMXBs with $P_{\rm
orb}\la 2$ days, most neutron star systems will be persistent X-ray
sources unless they have extreme mass ratios (i. e., $q<0.18P_{\rm
orb,hr}$). Consider the calculated results in Figs. 1-3, we conclude
that mass ejection is more likely to occur in wide systems with
light secondary (or in the late the evolution
of long-period LMXBs) via disruption of the accretion disk
inside the Roche lobe of the primary.

\begin{table}

\bc
\begin{minipage}[]{100mm}
\caption[]{Properties of  binary MSPs with measured masses}
\end{minipage}
\small
 \begin{tabular}{cccccr}
  \hline\noalign{\smallskip}
PSR Name & p (ms) & $P_{\rm orb,hr}$ (hr) &  $M_{p} (M_{\odot})$ & $M_2 (M_{\odot})$  & Ref.  \\
  \hline\noalign{\smallskip}
B1516+02B     & 7.95 & 164.6  & $2.08 \pm 0.19$         & $0.13^*$                  & 1 \\
B1855+09      & 5.36 & 295.2  & $1.6 \pm 0.2$           & $0.258^{+0.02}_{-0.03}$   & 2 \\
B1957+20      & 1.61 & 9.17   & $2.40 \pm 0.12$         & $0.035^{+0.01}_{-0.02}$   & 3 \\
J0024-7204H   & 3.21 & 60.9   & $1.41^{+0.04}_{-0.08}$  & $0.18^{+0.09}_{-0.02}$    & 4 \\
J0437-4715    & 5.76 & 137.8  & $1.76 \pm 0.20$         & $0.254 \pm 0.018$         & 5 \\
J0751+1807    & 3.48 & 6.3    & $1.26 \pm 0.14$         & $0.19 \pm 0.03$           & 6 \\
J1012+5307    & 5.26 & 14.5   & $1.68 \pm 0.22$         & $0.16 \pm 0.02$           & 7 \\
J1614-2230    & 3.15 & 208.8  & $1.97 \pm 0.04$         & $0.4^*$                   & 8 \\
J1713+0747    & 4.57 & 1627.8 & $1.53^{+0.08}_{-0.06}$  & $0.33 \pm 0.04$           & 9 \\
J1738+0333    & 5.85 & 8.52   & $1.55 \pm 0.55$         & $0.2 \pm 0.05$            & 10 \\
J1740-5340A   & 3.65 & 32.5   & $1.53 \pm 0.19$         & $0.22^*$                  & 11 \\
J1903+0327    & 2.15 & 2284.2 & $1.74 \pm 0.04$         & $1.05 \pm 0.02$           & 12 \\
J1909-3744    & 2.95 & 36.7   & $1.47^{+0.03}_{-0.02}$  & $0.2038 \pm 0.0022$       & 13 \\
J1911-5958A   & 3.27 & 20.6   & $1.4^{+0.16}_{-0.10}$   & $0.18 \pm 0.02$           & 14 \\
J1748-2446I   & 9.57 & 31.87  & $1.87^{+0.32}_{-0.07}$  & $0.24^*$                  & 15 \\
J1045-4509    & 7.47 & 97.92  & $<1.48$                 & $0.13$                    & 16 \\
J1804-2718    & 9.34 & 266.4  & $<1.73$                 & $0.2$                     & 16 \\
J2019+2425    & 3.93 & 1836   & $<1.51$                 & $0.32-0.35$               & 17 \\
J1023+0038    & 1.69 & 4.75   & $1.0-3.0$               & $0.14-0.42$               & 18 \\
J1738+0333    & 5.85 & 8.50   & $1.6 \pm 0.2$           & $0.2$                     & 19 \\
J0024-7204I   & 3.49 & 5.52   & $1.44$                  & $0.15$                    & 20 \\
J1518+0204B   & 7.95 & 164.64 & $2.08 \pm 0.19$         & $>0.13$                   & 21 \\
J1824-2452C   & 4.16 & 193.87 & $<1.367$                & $>0.26$                   & 21 \\
J0514-4002A   & 4.99 & 450.96 & $<1.52$                 & $>0.96$                   & 22 \\
J1748-2021I   & 9.57 & 31.87  & $1.3 \pm 0.02$          & $0.24$                    & 15 \\
  \noalign{\smallskip}\hline
\end{tabular}
\ec

\tablecomments{0.86\textwidth}{*: Medium companion mass from ATNF
pulsar database. 1.Freire et al 2008a,  2.Splaver 2004, 3.van
Kerkwijk et al. 2011, 4.Freire et al. 2003, Manchester et al. 1991,
5.Verbiest et al. 2008, 6.Nice et al. 2008, 7.Lange et al. 2001,
8.Demorest et al. 2010, 9.Splaver et al. 2005, 10.Freire et al.
2008b, 11.Kaluzny et al. 2003, 12.Champion et al. 2008, 13.Jacoby et
al. 2005, Hotan et al. 2006, 14.Bassa et al. 2006, 15.Ransom et al.
2005, 16.Thorsett \& Chakrabarty 1999, 17.Nice, Splaver, \& Stairs
2001, 18.Archibald et al. 2009, 19.Jacoby 2004, 20.Manchester et al.
1991, 21.Zhang et al. 2011, 22.Freire et al. 2007}

\label{Table1}

\end{table}

\section{Summary and discussion}

We can summarize the above results for the conditions of
pulsar-driven mass ejection as follows.

(1) Pulsar-driven mass ejection is likely to occur in wide LMXBs
(in case 1 evolution), in which the thermal instability in the accretion
disk can result in large variations in the accretion rate.

(2) LMXBs in case 2 evolution may barely be subject to mass ejection.

(3) If there is mass ejection,  it may be caused by the disruption
of the accretion disk leading to outflow through the $L_1$ point. When
the secondary is of extremely low mass, outflow from the $L_2$
point is also possible.

An important factor that was neglected in the above sections is
irradiation in LMXBs caused by accretion-generated X-rays. It was
realized that irradiation of the donor star or the accretion disk
can not only change the optical appearance of LMXBs (King \& Ritter
1999), but also their outburst properties (van Paradijs 1996), and
possibly the long-term evolution of the donor (Podsiadlowski 1991).
 Especially irradiation of the donor star can destabilize the mass
 transfer, and lead to mass transfer cycles  (Hameury et al. 1993), which
 differs drastically from the evolution we considered above:
 mass transfer is spasmodic with phases of high mass transfer driven
 by the thermal expansion of the convective envelope of the irradiated
 donor alternating with phases with low or no mass transfer, during which
 the donor readjusts towards thermal equilibrium of the un-irradiated
 star (B\"uning \& Ritter 2004). As pointed out by Ritter (2008),
 the effect of irradiation may be important in compact rather
 long-period LMXBs, since in the later the irradiation resulting from
 accretion is intermittent due to disk instability. However, even in
 the former systems,  the details of how irradiation of the donor
 influences
 the mass transfer process are very complicated, and have not
 been included in the calculations of the secular evolutions of
 I/LMXBs in a self-consistent way (Pfahl et al. 2003). It is then
 difficult to estimate the efficiency
 of mass ejection due to irradiation. On one hand, during the short ``high" state the
 mass transfer rate is enhanced and larger than the secular one,
 leading to shorter $p_{\rm eq}$, which is in favor of mass ejection
 (In Fig.~3, if the mass transfer rate is increased by a factor of $100$,
 both $p_{\rm eq}<p_{\rm cr}$ and $p_{\rm eq}<p'_{\rm cr}$ will be satisfied).
On the other hand, during the long ``low" state, the mass transfer rate
 becomes much lower than the secular one, so that less mass can be blown off.
 This is in contrast with the limit cycles due to disk instability, in which
 the accretion rate varies by a large factor but the mass transfer rate remains
 nearly unchanged.

 The current masses of MSPs may reveal possible evidence of
 mass accretion and ejection during  the previous LMXB evolution.
 In Table 1 we list the parameters of binary MSPs with measured
 masses\footnote{PSR J1903$+$0327 is a peculiar MSP with a main-sequence
 companion star, and is not considered  here.}.
 In general, long-period ($P_{\rm orb}\ga 20-30$ days) MSPs have masses
 around $1.4\,M_{\odot}$, suggesting that large amount of the
 transferred mass  was lost. In narrow
 systems ($P_{\rm orb}< 2-3$ days) the pulsar masses distribute from
 $\sim 1.3\,M_{\odot}$ to $\sim 2.4\,M_{\odot}$, indicating that
 both efficient mass accretion and mass ejection are possible, depending
 on the properties of the individual sources. However, a large fraction
 of them also have masses not far from $1.4\,M_{\odot}$. We speculate
 that irradiation-driven mass transfer cycles may help drive off the
 transferred mass from their companions in these systems.

Besides the standard recycling scenario, another possible way for
mass ejection in short-period LMXBs is
companion exchange. This should  happen only in the environment with
high stellar density like globular clusters. The incidence of black
widow pulsars in globular clusters is known to be far higher than in
the field. This leads King et al. (2003) to suggest that the MSPs in
dead wide binaries with white dwarf companions have exchanged them
for normal stars. Encounters and tides bring these new companions
into tight orbits. Due to intense accretion during the first mass
transfer phase, the neutron star's spin period may satisfy the
condition $p_{\rm eq}\le p_{\rm cr}$ or $p'_{\rm cr}$ with current
mass transfer rate, resulting in mass ejection from the binary.

Finally, the condition for mass ejection may be satisfied if the
accreting star in LMXBs is a strange star rather a neutron star. In
the former case the spin periods could be submilliseconds (e.g.,
Frieman \& Olinto 1989; Gourgoulhon et al. 1999; Possenti et al.
1999), and less than the critical periods given by Eqs.~(8) and
(14). These exotic objects might be formed by accretion-induced
collapse of white dwarfs in binary systems (Du et al. 2009).

\normalem
\begin{acknowledgements}

This work was supported by the Natural Science Foundation of China
under grant number 10873008 and the Ministry of Science and the
National Basic Research Program of China (973 Program 2009CB824800).

\end{acknowledgements}

\appendix

\section{Calculation of the disk pressure}

In the model of magnetic dominant disk Begelman and Pringle argued that
the amplified toroidal magnetic field pressure is $\sim \rho c_g v_k$.
They calculated the density inside the disc is
\begin{equation}
\rho = {{\Sigma} \over {2H}} \sim {{2} \over {3}} {{c^2} \over {GM\kappa}} \left( {c \over {c_g}} \right)^{3/2} {{\dot{M}} \over {\dot{M}_{Edd}}} {\alpha}^{-1} \left(R \over {R_g} \right)^{-9/4}
\end{equation}
here $\kappa$ is the opacity, $\dot{M}_{Edd}={4\pi GM}/{\kappa c}$
is the Eddington accretion rate and $R_g=GM/{c^2}$. In this model
the viscosity and scale height of the disc is given by $\nu = \alpha H v_A$
and $H/R \sim ({c_g} / {v_k})^{1/2}$ instead of $\nu = \alpha H c_g$ and
 $H/R = (c_g^2 + c_r^2)^{1/2} / v_k$ in SS model.
After these replacement then use the equation of angular momentum
\begin{equation}
\nu \Sigma = {{\dot{M}} \over {3\pi}} \left[ 1- {\left( {{R_*} \over R} \right)}^{1/2} \right]
\end{equation}
and energy
\begin{equation}
{{4\sigma} \over {3\tau}} {T_c}^4 = {{3GM\dot{M}} \over {8\pi R^3}} \left[ 1- {\left( {{R_*} \over R} \right)}^{1/2} \right]
\end{equation}
we can calculate the disk pressure is
\begin{equation}
P_{disk}= {\alpha}^{-17/18} {(GM)^{61/36} \over {6\pi c^2}} \left({3\kappa} \over {32{\pi}^2\sigma}\right)^{-1/18}
{{\dot{M}^{8/9}} \over {R_n}} \left( {k \over {\mu m_p}}\right)^{-2/9} R^{-91/36} \left[1-\left( {R_* \over R} \right)^{1/2} \right]^{-1/9}.
\end{equation}

\label{lastpage}


\begin{thebibliography}{99}
\small \setlength{\itemindent}{-3mm} \setlength{\itemsep}{-0.5mm}
\setlength{\baselineskip}{4.5mm}



\bibitem[{Archibald}{et~al.~}(2009)]{Asr+09} Archibald A. M., Stairs I. H., Ransom S. M. et al., 2009, Science, 324, 1411

\bibitem[{Balbus \& Hawley}(1998)]{Bh98} Balbus S. A., Hawley J. F., 1998, Rev. Mod. Phys., 70, 1

\bibitem[{Bassa}{et~al.~}(2006)]{Bvk+06} Bassa, C.G., van Kerkwijk, M.H., Koester,  D. \& Verbunt, F. 2006, \aap , 456, 295

\bibitem[{Begelman \& Pringle}(2007)]{Bp07} Begelman, M. C. \& Pringle, J. E., 2007, \mnras , 375, 1070

\bibitem[{Bhattacharya \& van den Heuvel}(1991)]{Bv91} Bhattacharya, D., van den Heuvel, E. P. J., 1991, Phys. Rep., 203, 1

\bibitem[{B\"uning \& Ritter}(2004)]{BR04} B\"uning, A, \& Ritter, H., 2004, A\&A 423, 281

\bibitem[{Burderi}{et~al.~}(2002)]{Bdb02} Burderi, L., D'Antona, F. \& Burgay, M. 2002, \apj , 574, 325

\bibitem[{Burderi}{et~al.~}(1999)]{Bpc+99} Burderi, L., Possenti, A., Colpi, M., Di Salvo,T., \& D`Amico, N. 1999, \apj , 519, 285

\bibitem[{Burderi}{et~al.~}(2001)]{Bpd+01} Burderi, L., Possenti, A., D'Antona, F. , et al. 2001, \apj , 560, L71

\bibitem[{Champion}{et~al.~}(2008)]{Crl+08} Champion, D.J., Ransom, S.M., Lazarus, P. et al., 2008, Science, 320, 1309

\bibitem[{D'Amico et al.}(2001)]{DA01} D'Amico, N., Possenti, A., Manchester, R. N., Sarkissian, J., Lyne, A. G., Camilo, F., 2001, \apj, 561, L89

\bibitem[{Davies \& Hansen}(1998)]{Dh98} Davies, M.B. \& Hansen, B.M., 1998, \mnras , 301, 15

\bibitem[{Demorest}{et~al.~}(2010)]{Dpr10} Demorest P. B., Pennucci T., Ransom, S. M., et al. 2010, Nature, 467, 1081

\bibitem[{Du et al.}(2009)]{Du09} Du, Y.-J., Xu, R.-X., Qiao, G.-J., \& Han, J.-L. 2009, MNRAS, 399,
1587

\bibitem[{Ferraro}{et~al.~}(2001)]{F01} Ferraro, F. R., Possenti, A., D'Amico, N., Sabbi, E. 2001, \apj , 561, L93

\bibitem[{Frank, King,  \& Raine}(2002)]{FKR02} Frank, J., King, A. R., \& Raine, D. 2002, Accretion power in astrophysics, Cambridge University Press

\bibitem[{Freire}{et~al.~}(2008)]{Fwv08} Freire, P.C.C., Wolszcan, A., van den Berg,M. et al, 2008a, \apj , 679, 1433

\bibitem[{Freire}{et~al.~}(2007)]{Frg07} Freire P. C., Ransom S. M. \& Gupta Y. 2007, \apj , 662, 1177

\bibitem[{Freire}{et~al.~}(2003)]{Fck03} Freire, P.C.C., Camilo, F., Kramer, M., et al. 2003 \mnras , 340, 1359

\bibitem[{Freire}{et~al.~}(2008)]{Fjb08} Freire, P.C.C., Jacoby, B.A., \& Bailes,M., 2008b, AIP Conference Proceedings 983, 488; arXiv:0711.1880

\bibitem[{Frieman \& Olinto}(1989)]{FO89} Frieman, J. A., \& Olinto, A. V. 1989, Nature, 341, 633

\bibitem[{Fruchter, Stinebring, \& Taylor}(1988)]{F88} Fruchter, A. S., Stinebring, D. R., Taylor, J. H. 1988, \nat , 312, 255

\bibitem[{Ghosh \& Lamb}(1979)]{Gl79} Ghosh, P., \& Lamb, F. K. 1979, \apj , 234, 296

\bibitem[{Gourgoulhon et al.}(1999)]{G99} Gourgoulhon, E., Haensel, P., Livine, R., Paluch, E., Bonazzola, S.,
\& Marck, J.-A. 1999, \aap, 349, 851

\bibitem[{Hakala}{et~al.~}(1999)]{Hmd99} Hakala P. J., Muhli P., Dubus G., 1999, \mnras , 306, 701

\bibitem[{Hameury}{et~al.~}(1993)]{Hkl+93} Hameury J. M., King A. R., Lasota J.P. et al. 1993, \aap , 277, 81

\bibitem[{Hotan}{et~al.~}(2006)]{Hbo06} Hotan, A.W., Bailes, M. \& Ord, S.M., 2006, \mnras , 369, 1502

\bibitem[{Illarionov \& Sunyaev}(1975)]{Is75} Illarionov, A. F. \& Sunyaev, R. A. 1975, \aap , 39, 185

\bibitem[{Jacoby}(2004)]{Jacoby04} Jacoby B. A. 2004, PhD thesis, Calfornia Institute of Technology

\bibitem[{Jacoby}{et~al.~}(2005)]{Jhb+05} Jacoby, B. A., Hotan, A., Bailes, M., Ord, S., \& Kulkarni, S. R. 2005, \apj , 629, L113

\bibitem[{Kaluzny}{et~al.~}(2003)]{Krt03} Kaluzny, J., Rucinski, S. M., \& Thompson, I. B. 2003, \aj , 125, 1546

\bibitem[{Kiziltan}{et~al.~}(2010)]{Kkt10} Kiziltan, B., Kottas, A., \& Thorsett, S. E. 2010, (arXiv:1011.4291)

\bibitem[{King}(1988)]{King88} King, A. R. 1988, QJRAS, 29, 1

\bibitem[{King}{et~al.~}(2003)]{Kdb03} King, A.R., Davies, M.B. \& Beer, M.E., 2003, \mnras , 345, 678

\bibitem[{King}{et~al.~}(1996)]{Kkb96} King, A. R., Kolb, U., Burderi, L., 1996, \apj , 464, L127

\bibitem[{King \& Ritter}(1996)]{Kr99} King, A. R. \& Ritter, H.
1999, MNRAS, 309, 253

\bibitem[{Landau \& Lifschitz}(1958)]{Ll58} Landau, L., \& Lifschitz, E. 1958, Classical Theory of Fields (Elmsford: Pergamon)

\bibitem[{Lange}{et~al.~}(2001)]{Lcw01} Lange, Ch., Camilo, F., Wex, N., et al. 2001 \mnras , 326, 274 (2001)

\bibitem[{Lasota}(2001)]{La01} Lasota, J.-P. 2001, NewAR, 45, 449

\bibitem[{Li}{et~al.~}(2010)]{Lml10} Li, S.-L., Miller, N. Lin, D.N.C., Fortney, et al. 2010, Nature, 463, 1054

\bibitem[{Manchester}{et~al.~}(1991)]{Mlr+91} Manchester R. N., Lyne A. G., Robinson C., et al. 1991, Nature, 352, 219

\bibitem[{Nice}{et~al.~}(2001)]{Nss+01} Nice, D. J., Splaver, E. M. \& Stairs, I. H. 2001, \apj , 549, 516

\bibitem[{Nice}{et~al.~}(2005)]{Nss+05} Nice, D. J., Splaver, E. M., Stairs, I. H., et al. 2005, \apj , 634, 1242

\bibitem[{Nice}{et~al.~}(2008)]{Nsk08} Nice, D.J., Stairs, I.H. \& Kasian, L.E., 2008, AIP Conference Proceedings 983, 453

\bibitem[{Pfahl}{et~al.~}(2003)]{Prp03} Pfahl, E. D., Rappaport, S. \& Podsiadlowski, P., 2003, \apj , 597 1036

\bibitem[{Phillips \& Podsiadlowski}(2002)]{PP02} Phillips, S. N. \& Podsiadlowski, P. 2002, MNRAS, 337, 431

\bibitem[{Podsiadlowski}(1991)]{Podsiadlowski91} Podsiadlowski, P. 1991, Nature, 350, 136

\bibitem[{Podsiadlowski}{et~al.~}(2002)]{Prp02} Podsiadlowski, P., Rappaport, S. \& Pfahl, E. D. 2002 \apj , 565, 1107

\bibitem[{Possenti}{et~al.~}(1999)]{P99} Possenti, A., Colpi, M., Geppert, U., Burderi, L., \& D'Amico, N. 1999, ApJS, 125, 463

\bibitem[{Ransom}{et~al.~}(2005)]{Rhs05} Ransom, S. M., Hessels, J.W.T., Stairs, I.H., et al. 2005, Science, 307, 892

\bibitem[{Ritter}(2008)]{Ritter08} Ritter, H. 2008, NewAR, 51, 869

\bibitem[{Ruderman}{et~al.~}(1989)]{Rst89} Ruderman, M., Shaham, J. \& Tavani, M. 1989, \apj , 336, 507

\bibitem[{Shafter \& Misselt}(2006)]{Sm06} Shafter A.W., Misselt K. A., 2006, \apj , 644, 1104

\bibitem[{Shaham \& Tavani}(1991)]{St90} Shaham, J. \& Tavani, M. 1991, \apj , 377, 588

\bibitem[{Shakura \& Sunyaev}(1973)]{Ss73} Shakura, N. I. \& Sunyaev, R. A., 1973, \aap , 24, 337

\bibitem[{Splaver}(2004)]{Splaver04} Splaver, E. M., 2004, PhD Thesis, Princeton University

\bibitem[{Splaver}{et~al.~}(2005)]{Sns+05} Splaver, E. M., Nice, D. J., Stairs, I. H., Lommen, A. N., \& Backer, D. C. 2005, \apj , 620, 405

\bibitem[{Tauris et al.}(2011)]{TLK11} Tauris, T. M., Langer, N., \& Kramer, M. 2011, \mnras, in press
(arXiv:1103.4996)

\bibitem[{Tauris \& Savonije}(1999)]{Ts99} Tauris, T. M. \& Savonije, G. J. 1999, \aap , 350, 928

\bibitem[{Thorsett \& Chakrabarty}(1999)]{Tc99} Thorsett, S. E. \& Chakrabarty, D. 1999, \apj , 512, 288

\bibitem[{van den Heuvel \& van Paradijs}(1988)]{HP88} van den Heuvel, E. P. J. \& van Paradijs, J. 1988, \nat, 334, 227

\bibitem[{van Kerkwijk}{et~al.~}(2011)]{Vbk10} van Kerkwijk, M. H., Breton, R., \& Kulkarni, S. R., 2011, \apj , 728, 95

\bibitem[{van Paradijs}(1996)]{P96} van Paradijs, J., 1996, ApJ 464, L139

\bibitem[{Verbiest}{et~al.~}(2008)]{Vbv+08} Verbiest, J. P. W., Bailes, M., van Straten, W. et al., 2008, \apj , 679, 675

\bibitem[{Verbunt \& Zwaan}(1981)]{Vz81} Verbunt, F., \& Zwaan, C. 1981, \aap , 100, L7

\bibitem[{Webbink}{et~al.~}(1983)]{Wrs83} Webbink, R. F., Rappaport, S., \& Savonije, G. J. 1983, \apj , 270, 678

\bibitem[{Zhang}{et~al.~}(2011)]{Zwz+11} Zhang C. M., Wang J., Zhao Y. H. et al. 2011, \aap , 527, 83

\end{thebibliography}
\end{document}